\documentclass[fleqn,aps,pra,amsmath,amssymb,preprint,showpacs]{revtex4-1}

\usepackage{float}
\usepackage{graphicx}
\usepackage{textcomp}
\usepackage{braket}
\usepackage{enumerate}

\renewcommand\Im{\operatorname{Im}}

\begin{document}

\title{Wavelength mismatch effect in electromagnetically induced absorption}

 \author{Vineet Bharti}
 \affiliation{Department of Physics, Indian Institute of
 Science, Bangalore 560\,012, India}

 \author{Ajay Wasan}
 \altaffiliation{Department of Physics, Indian Institute of Technology, Roorkee 247\,667, India}

\author{Vasant Natarajan}
 \affiliation{Department of Physics, Indian Institute of
 Science, Bangalore 560\,012, India}
 \email{vasant@physics.iisc.ernet.in}
 \homepage{www.physics.iisc.ernet.in/~vasant}

\begin{abstract}
We present a theoretical investigation of the phenomenon of electromagnetically induced absorption (EIA) in a 4-level system consisting of vee and ladder subsystems. The four levels are coupled using one weak probe field, and two strong control fields. We consider an experimental realization using energy levels of Rb. This necessitates dealing with different conditions of wavelength mismatch---near-perfect match where all three wavelengths are approximately equal; partial mismatch where the wavelength of one control field is less than the other fields; and complete mismatch where all three wavelengths are unequal. We present probe absorption profiles with Doppler averaging at room temperature to account for experiments in a room temperature Rb vapor cell. Our analysis shows that EIA resonances can be studied using Rydberg states excited with diode lasers.\\

\noindent
\textbf{Keywords}: Electromagnetically induced absorption, wavelength mismatch effect, Coherent control.
\end{abstract}

\maketitle

\section{Introduction}
The phenomenon of electromagnetically induced transparency (EIT) is one where an initially absorbing medium is rendered transparent by having a strong control field on an auxiliary transition. EIT therefore requires the presence of at least three levels, and consequently has been studied in the three canonical types of 3-level systems, namely lambda ($ \Lambda $), vee (V), and ladder ($ \Xi $) \cite{HAR97,FIM05}.  The presence of additional levels allows the use of additional control fields, which leads to a modification of the EIT window. Such extended systems allow the possibility of electromagnetically induced absorption (EIA), a phenomenon where an initially absorbing medium shows enhanced absorption at line center. EIA has been studied---both theoretically and experimentally---primary in N-type ($ \Lambda + $V) 4-level systems \cite{GWR04,CPN12,BMW09,BHW13}. We have recently shown that EIA resonances are also possible in a new kind of 4-level system made by adding vee and ladder configurations \cite{BHN15}. This configuration opens up the possibility of observation of EIA in Rydberg atoms, and has potential applications in switching a medium between sub-luminal and super-luminal light propagation. 

When EIT and EIA phenomena are studied in an atomic vapor at a finite temperature, the thermal motion affects the probe absorption profile due to the Doppler effect. The Doppler shift is determined by the velocity of the atom and the wavelengths of the fields, which then determines the detunings $ \Delta_p $ and $ \Delta_c $ in the atom's frame of reference. Since the two-photon absorption for a V-type system contains the term $ \Delta_p - \Delta_c $, the medium can be made Doppler free by choosing the beams to be co-propagating and of the same wavelengths. Similarly, the two-photon absorption for a ladder-type system contains the term $ \Delta_p + \Delta_c $, and hence the medium can be made Doppler free by using counter-propagating beams that are of the same wavelength. This corresponds to a perfect wavelength matching condition, and results in a narrowing of the transparency window \cite{BZM99,IKN08,BHW12}. 

However, experimental realization in a real atomic system cannot always satisfy this perfect wavelength matching condition. The difference in wavelengths then leads to a modification of the absorption profile. EIT, under conditions of wavelength mismatch, has therefore been investigated in the three types of 3-level systems \cite{BZM99,SFD96,BZF98,VBA07,KUS09,UCR13}. It has also been studied in a Y-type 4-level system which only shows EIT (and not EIA) \cite{MIS12,BHW14}.

In this work, we extend the study of wavelength mismatch to the phenomenon of EIA in a 4-level system made using vee and ladder subsystems. This analysis is particularly important when the uppermost state is a Rydberg state. In fact, one proposed method of populating Rydberg states using diode lasers \cite{FAN11} results in the EIA window disappearing due to Doppler averaging of the mismatched wavelengths. We therefore propose an alternate scheme for excitation of Rydberg state.

\section{Theoretical considerations}

\begin{figure}
 	\centering
 	\includegraphics[width=.3\textwidth]{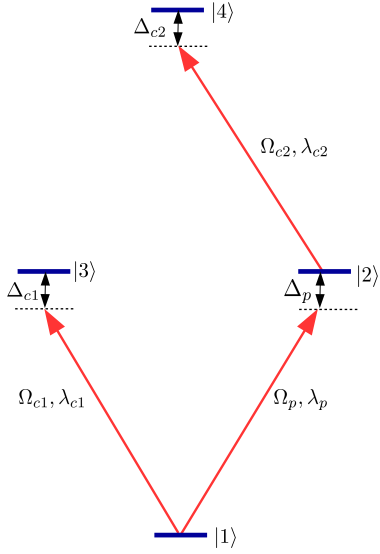}
 	\caption{(Color online) 4-level vee + ladder system under consideration.}
 	\label{levels}
\end{figure}

The combination of vee and ladder configuration used to form the 4-level system is shown in Fig.\ \ref{levels}. The weak probe field is common to both configurations, and couples levels $\ket{1}$ and $\ket{2}$. The vee configuration is formed using level $\ket{3}$, with a strong control field coupling levels $\ket{1}$ and $\ket{3}$. The ladder configuration is formed using level $\ket{4}$, with a strong control field coupling levels $\ket{2}$ and $\ket{4}$. The respective fields are denoted by: wavelength $\lambda $, strength $\Omega $, and detuning $ \Delta $. 
 
The Hamiltonian of the system (in the rotating wave approximation) for an atom moving with velocity $v$ is given by 
\begin{equation}
\begin{aligned}
H &= \dfrac{\hbar}{2} \left[ \, \Omega_{p} \ket{1} \bra{2}+\Omega_{c1} \ket{1} \bra{3}+\Omega_{c2} \ket{2} \bra{4} \, \right]+ {\rm h.c.} \\
&+\hbar \left[ \, \left( \Delta_{p} \pm k_{p}v \right)\ket{2} \bra{2}+ \left( \Delta_{c1} \pm k_{c1}v \right)\ket{3} \bra{3} \right. \\
&+ \left. \left( \Delta_{p}\pm k_{p}v+\Delta_{c2}\mp k_{c2}v \right) \ket{4} \bra{4} \, \right]
\label{Hamiltonion}
\end{aligned}
\end{equation}
where $ k = 2\pi/\lambda $, is the wavevector of the respective field, and $kv$ is the corresponding Doppler shift seen by the atom. As discussed earlier, the Doppler shift is minimized by having the probe and control 1 fields co-propagating (for the vee subsystem), and control 2 field counter-propagating (for the ladder subsystem). The equations of motion for the various density-matrix elements for this system are given in Appendix 1.

The observable in this work is the absorption of the probe field in the presence of the two control fields. It is proportional to the imaginary part of the coherence between levels $\ket{1}$ and $ \ket{2}$, and is given by $ \Im\{{\rho_{12}\Gamma_2/\Omega_p}\}$. As shown in Appendix 2, the system reaches steady state after a few lifetimes of $\ket{2}$, which means it is reached in 2 \textmu s or less. For comparison to earlier expressions for EIT in the three-level subsystems \cite{KPW05,DPW06}---which can be obtained by setting the appropriate control Rabi frequency to $0$---we give below an analytic expression for $\rho_{12}$ in steady state and to first order in $ \Omega_p$.
\begin{equation}
\begin{aligned}
\rho_{12}&=\displaystyle{-\frac{i\Omega_{p}\rho_{11}}{2\gamma_{12}\beta}+
\frac{i\Omega_{p}|\Omega_{c1}|^2(\rho_{11}-\rho_{33})}{8\gamma_{12}\gamma_{31}\gamma_{32}\beta}
\left(1-\frac{|\Omega_{c2}|^2}{4\gamma_{32}\gamma_{34}\alpha}-
\frac{|\Omega_{c2}|^2}{4\gamma_{14}\gamma_{34}\alpha}\right)}
\label{solution}
\end{aligned}
\end{equation}
where
\begin{equation*}
\begin{aligned}
&\rho_{11}-\rho_{33} = \left[ 1 + \frac{|\Omega_{c1}|^2}{2\left(\frac{\Gamma_{3}^2}{4}+(\Delta_{c1}\pm k_{c1}v)^2\right)} \right]^{-1} \,\\
& \beta=1+\displaystyle{\frac{|\Omega_{c1}|^2}{4\gamma_{12}\gamma_{32}}+\frac{|\Omega_{c2}|^2}{4\gamma_{12}\gamma_{14}}
-\frac{|\Omega_{c1}|^2|\Omega_{c2}|^2}{16\gamma_{12}\gamma_{34}\alpha}\left[\frac{1}{\gamma_{32}}+\frac{1}{\gamma_{14}}\right]^2} \,\\ 
&\alpha=\displaystyle{1+\frac{|\Omega_{c1}|^2}{4\gamma_{14}\gamma_{34}}+\frac{|\Omega_{c2}|^2}{4\gamma_{32}\gamma_{34}}} \,\\ 
& \gamma_{12}=\displaystyle{\left(-\frac{\Gamma_{2}}{2}+i(\Delta_{p}\pm k_{p}v)\right)} \,\\
& \gamma_{31}=\displaystyle{\left(-\frac{\Gamma_{3}}{2}-i(\Delta_{c1}\pm k_{c1}v)\right)} \,\\
& \gamma_{14}=\displaystyle{\left(-\frac{\Gamma_{4}}{2}+i(\Delta_{p}\pm k_{p}v+\Delta_{c2}\mp k_{c2}v)\right)} \,\\ 
& \gamma_{32}=\displaystyle{\left(-\frac{\Gamma_{2}+\Gamma_{3}}{2}-i(\Delta_{c1}\pm k_{c1}v-\Delta_{p}\mp k_{p}v)\right)} \,\\ 
& \gamma_{34}=\displaystyle{\left(-\frac{\Gamma_{3}+\Gamma_{4}}{2}+i(\Delta_{p}\pm k_{p}v+\Delta_{c2}\mp k_{c2}v-\Delta_{c1}\mp k_{c1}v)\right)} \,
\end{aligned}
\end{equation*}
In Eq.\ \eqref{solution}, $\Gamma$'s are the relaxation rates of corresponding energy levels.

\section{Results and discussion}

We will analyze different mismatching conditions by looking at experimental realization of the 4 levels using appropriate energy levels of Rb. Probe absorption given by Eq.\ \eqref{solution} is shown for the various cases, along with Doppler averaging to account for the situation in room temperature vapor. In all cases, level $\ket{1}$ is taken to be the $\rm 5 \, S_{1/2}$ ground state with $ \Gamma = 0 $. The results are shown for resonant control fields, i.e.\ $ \Delta_{c1} = \Delta_{c2} = 0 $. The strengths of both fields are taken to be equal to $ 3 \Gamma_2 $, i.e.\ $ \Omega_{c1} = \Omega_{c2} = 3 \Gamma_2 $. The strength of the probe field is taken to be small enough to satisfy the weak probe condition, i.e. $\Omega_p/2\pi = 0.001 $ MHz. 

\subsection{Near-perfect match}

This situation corresponds to $ \lambda_p \approx \lambda_{c1} \approx \lambda_{c2} $. It is realized using the following states:

\begin{center}
\begin{tabular}{c c c c}
\hline \hline
Level & Rb state & $\Gamma/2\pi$ & Wavelength \\ 
& & (MHz) & (nm) \\
\hline \\[-0.5em]
$\ket {2}$ & $\rm 5P_{3/2}$ & $6.1$ & $ \lambda_p = 780 $  \vspace*{0.5em}  \\
$\ket {3}$ & $\rm 5P_{1/2}$ & $5.9$ & $ \lambda_{c1} = 795 $  \vspace*{0.5em} \\  
$\ket {4}$ & $\rm 5D_{5/2}$ & $0.68$ & $ \lambda_{c2} = 776 $ \vspace*{0.5em} \\
\hline \hline  
\end{tabular}
\end{center}

The results for this case are shown in Fig.\ 2. The 3D contour plot in part (a) shows the variation with both atomic velocity and probe detuning. There is an EIA peak on resonance and increased absorption at $\Delta_p = \pm 3 \Gamma_2$, because the poles in Eq.\ \eqref{solution} are at ($ \pm \Omega_{c1} \pm \Omega_{c2})/2 $. The near-equal wavelengths makes the geometry of the beams Doppler free. Therefore, the EIA peak at line center survives Doppler averaging. These results are shown in part (b), and correspond to a Maxwell-Boltzmann distribution covering a velocity range of $ -500 $ to $ +500 $ m/s, which is adequate for room temperature vapor.

\begin{figure}
	\centering
	\includegraphics[width=0.85\textwidth]{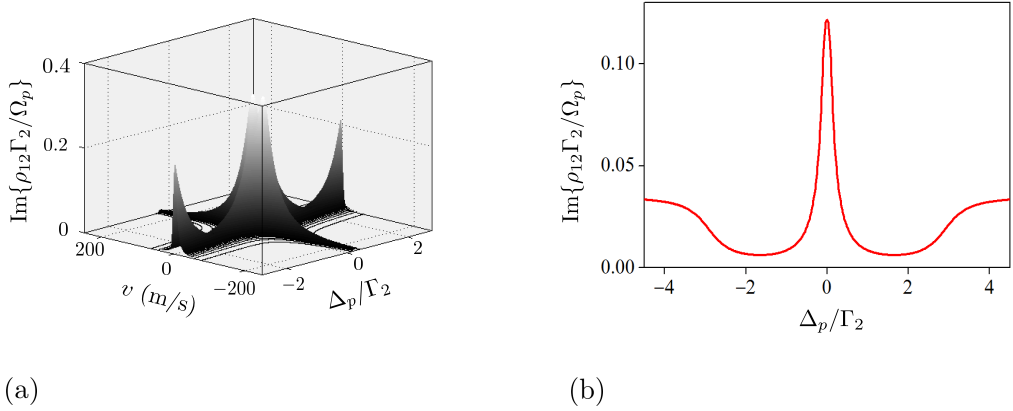}
	\caption{(Color online) Probe absorption given by $\Im\{{\rho_{12}\Gamma_2/\Omega_p}\}$ for the case of near-perfect match. (a) Contour plot showing variation with both probe detuning and atomic velocity. (b) Variation with probe detuning after Doppler averaging at room temperature.}
	\label{near_match}
\end{figure}

\subsection{Partial mismatch}

This situation corresponds to the wavelength of the probe field being approximately equal to that of one control field, but differing from the other one. This is further distinguished into two, depending on the relative value of the mismatched wavelength.

\subsection*{Case 1. $\lambda_p \approx \lambda_{c1} > \lambda_{c2}$}

This case is realized using the following states:
\begin{center}
\begin{tabular}{c c c c}
\hline \hline
Level & Rb state & $\Gamma/2\pi$ & Wavelength  \\ 
& & (MHz) & (nm) \\
\hline \\[-0.5em]
$\ket {2}$ & $\rm 5P_{3/2}$ & $ 6.1 $ & $ \lambda_p = 780$  \vspace*{0.5em}  \\
$\ket {3}$ & $\rm 5P_{1/2}$ & $ 5.9 $ & $ \lambda_{c1} = 795$ \vspace*{0.5em} \\  
$\ket {4}$ & \parbox{4em}{\centering $\rm 44D_{5/2}$\\(Rydberg)} & $ 0.3 $ & $ \lambda_{c2} = 480 $ \vspace*{0.5em} \\
\hline \hline  
\end{tabular}
\end{center}

\begin{figure}[tbh!]
	\centering
	\includegraphics[width=0.85\textwidth]{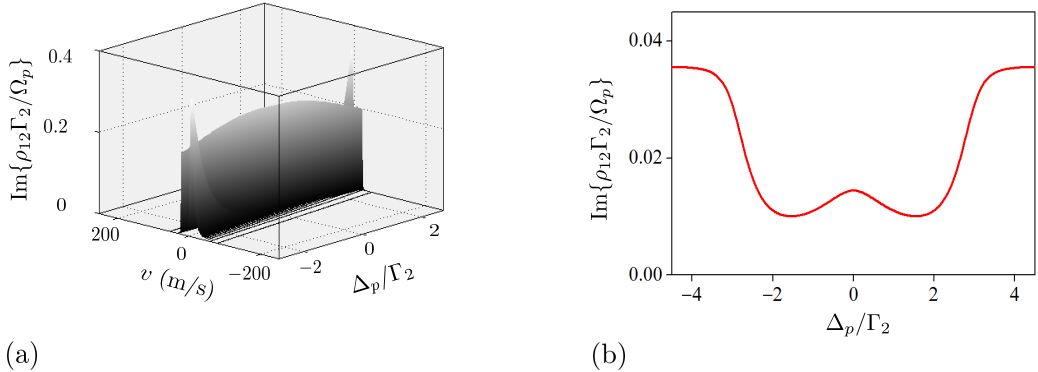}
\caption{(Color online) Probe absorption given by $\Im\{{\rho_{12}\Gamma_2/\Omega_p}\}$ for case 1 of partial mismatch. (a) Contour plot showing variation with both probe detuning and atomic velocity. (b) Variation with probe detuning after Doppler averaging at room temperature.}
	\label{partial_mismatch_case1}
\end{figure}

The results for this case are shown in Fig.\ \ref{partial_mismatch_case1}. Because $ \lambda_p > \lambda_{c2} $, the two photon resonance for the ladder subsystem is not Doppler free, i.e.\ $k_p v - k_{c2}v \neq 0 $. This results in probe absorption spreading over different velocity classes, which causes the Doppler-averaged EIA resonance---shown in part (b)---to remain as a peak even though it is less prominent than for zero-velocity atoms. This is important when studying EIA with Rydberg states, since all wavelengths---780, 795, and 480---are accessible with diode laser systems.

\subsection*{Case 2. $\lambda_p \approx \lambda_{c2} > \lambda_{c1}$}

This case realized using the following states:
\begin{center}
\begin{tabular}{c c c c}
\hline \hline
Level & Rb state & $\Gamma/2\pi$ & Wavelength  \\ 
& & (MHz) & (nm) \\
\hline \\[-0.5em]
$\ket {2}$ & $\rm 5P_{3/2}$ & $ 6.1 $ & $ \lambda_p = 780 $  \vspace*{0.5em}  \\
$\ket {3}$ & $\rm 6P_{1/2}$ & $ 1.3 $ & $ \lambda_{c1} = 420 $  \vspace*{0.5em} \\  
$\ket {4}$ & $\rm 5D_{5/2}$ & $ 0.68 $ & $ \lambda_{c2} = 776 $ \vspace*{0.5em} \\
\hline \hline  
\end{tabular}
\end{center}

\begin{figure}[tbh!]
	\centering
	\includegraphics[width=0.85\textwidth]{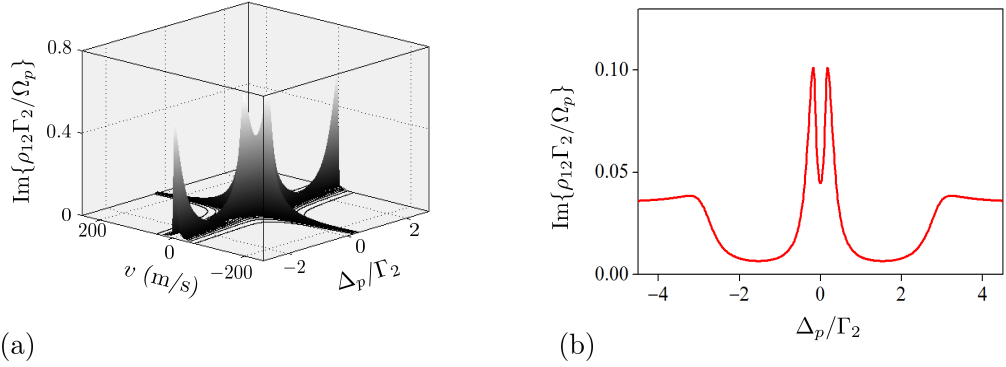}
\caption{(Color online) Probe absorption given by $\Im\{{\rho_{12}\Gamma_2/\Omega_p}\}$ for case 2 of partial mismatch. (a) Contour plot showing variation with both probe detuning and atomic velocity. (b) Variation with probe detuning after Doppler averaging at room temperature.}
	\label{partial_mismatch_case2}
\end{figure}

The curves obtained in this case are shown in Fig.\ \ref{partial_mismatch_case2}. The near-equal wavelengths of the probe and control 2 fields makes the two photon absorption for the ladder subsystem nearly Doppler free, hence absorption does not spread along different velocity classes. This causes the Doppler-averaged EIA peak---shown in part (b)---to split into two.

\subsection{Complete mismatch}

In this regime, the wavelength of all three fields are unequal, i.e.\ $\lambda_{p} \neq \lambda_{c1} \neq \lambda_{c2}$. This regime is also subdivided into two depending on the relative magnitude of the probe wavelength.

\subsection*{Case 1. $\lambda_p > ( \lambda_{c1} $ \rm{and} $\lambda_{c2} )$}

This case is realized using the following states:

\begin{center}
\begin{tabular}{c c c c}
\hline \hline
Level & Rb state & $\Gamma/2\pi$ & Wavelength  \\ 
& & (MHz) & (nm) \\
\hline \\[-0.5em]
$\ket {2}$ & $\rm 5P_{3/2}$ & $ 6.1 $ & $ \lambda_p = 780 $  \vspace*{0.5em}  \\
$\ket {3}$ & $\rm 6P_{1/2}$ & $ 1.3 $ & $ \lambda_{c1} = 420 $ \vspace*{0.5em} \\  
$\ket {4}$ & \parbox{4em}{\centering $\rm 44D_{5/2}$\\(Rydberg)} & $ 0.3 $ & $ \lambda_{c2} = 480 $ \vspace*{0.5em} \\
\hline \hline  
\end{tabular}
\end{center}

\begin{figure}[tbh!]
	\centering
	\includegraphics[width=0.85\textwidth]{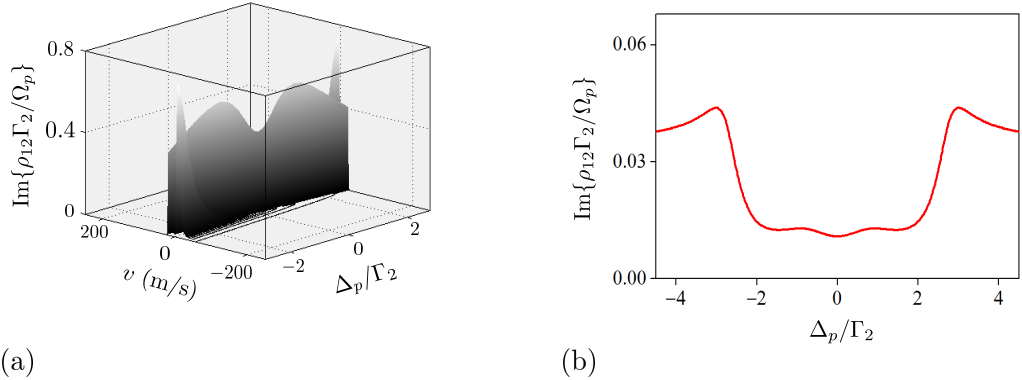}
\caption{(Color online) Probe absorption given by $\Im\{{\rho_{12}\Gamma_2/\Omega_p}\}$ for case 1 of complete mismatch. (a) Contour plot showing variation with both probe detuning and atomic velocity. (b) Variation with probe detuning after Doppler averaging at room temperature.}
	\label{complete_mismatch_case1}
\end{figure}

In this regime, probe absorption, which is by definition not Doppler free, both spreads along velocity classes and splits on resonance. The results are shown in Fig.\ \ref{complete_mismatch_case1}, and are in effect a convolution of the two cases considered previously. As can be seen from the Doppler-averaged spectrum shown in part (b), the transparency window at line center becomes very broad.

\subsection*{Case 2. $\lambda_p < ( \lambda_{c1} $ \rm{and} $\lambda_{c2} )$} 

This is an inverted situation for the vee and ladder subsystems. It is realized using the following states:

\begin{center}
\begin{tabular}{c c c c}
\hline \hline
Level & Rb state & $\Gamma/2\pi$ & Wavelength  \\ 
& & (MHz) & (nm) \\
\hline \\[-0.5em]
$\ket {2}$ & $\rm 6P_{3/2}$ & $ 1.4 $ & $ \lambda_p = 420 $  \vspace*{0.5em}  \\
$\ket {3}$ & $\rm 5P_{1/2}$ & $ 5.9 $ & $ \lambda_{c1} = 795 $ \vspace*{0.5em} \\  
$\ket {4}$ & \parbox{4em}{\centering $\rm 44D_{5/2}$\\(Rydberg)} & $ 0.3 $ & $ \lambda_{c2} = 1016 $ \vspace*{0.5em} \\
\hline \hline  
\end{tabular}
\end{center}

\begin{figure}[tbh!]
	\centering
	\includegraphics[width=0.85\textwidth]{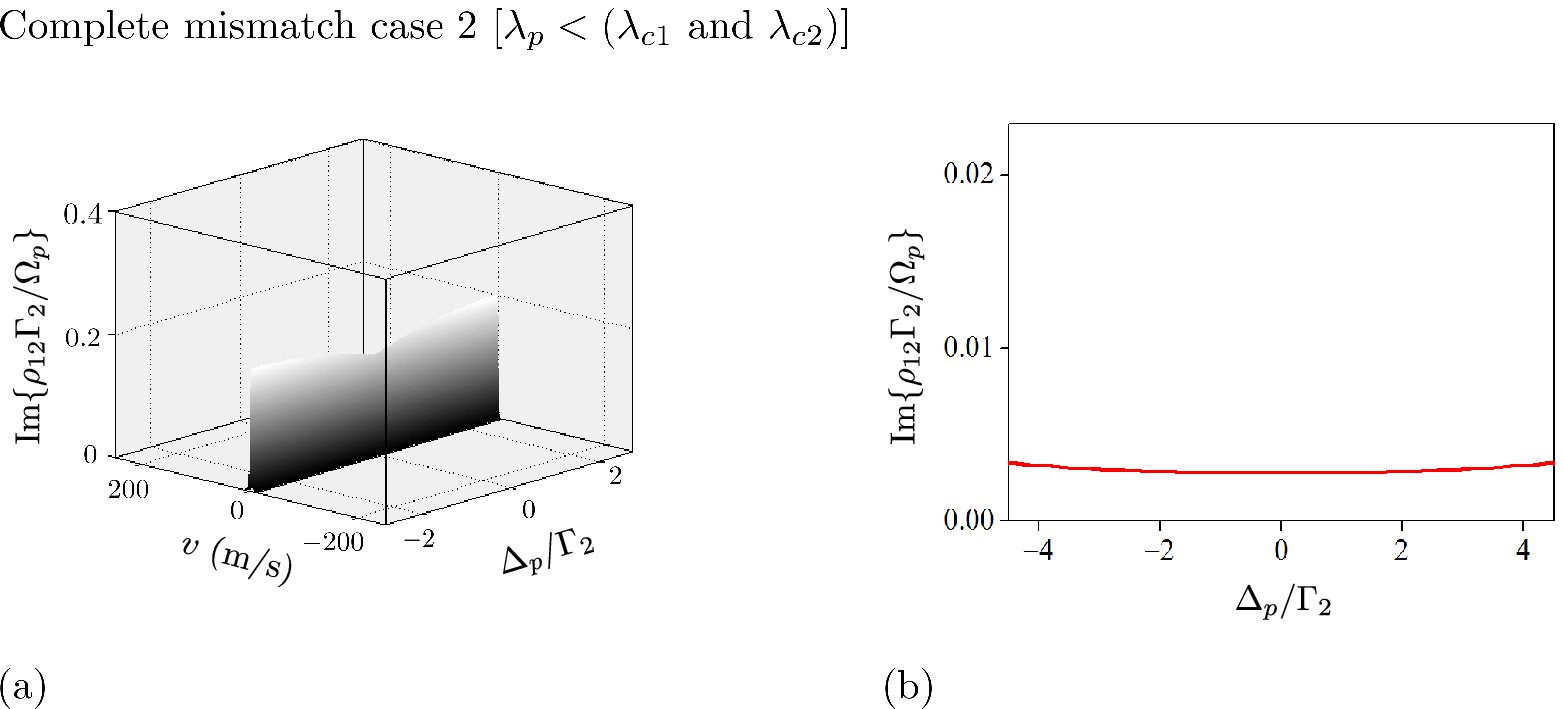}
\caption{(Color online) Probe absorption given by $\Im\{{\rho_{12}\Gamma_2/\Omega_p}\}$ for case 2 of complete mismatch. (a) Contour plot showing variation with both probe detuning and atomic velocity. (b) Variation with probe detuning after Doppler averaging at room temperature.}
	\label{complete_mismatch_case2}
\end{figure}

In this case, both EIT and EIA features are not observed, because the Autler-Townes doublet of the two 3-level subsystems begin to overlap \cite{BZM99}. The results, shown in Fig.\ \ref{complete_mismatch_case2}, bear out this expectation. This prediction also makes the proposal for accessing Rydberg states in Rb using diode lasers in Ref.\ \cite{FAN11} useless for observing EIA. 

\section{Conclusion}
In summary, we have theoretically studied EIA resonances in a vee + ladder 4-level system. We have considered implementation of the level scheme with relevant energy levels of Rb, which necessitates the dealing of different regimes of wavelength mismatch. We present 3D contour plots showing variation of probe absorption with both probe detuning and atomic velocity. The detailed behavior in each case can be qualitatively understood based on the expression for probe absorption. We further present results after Doppler averaging at room temperature, which corresponds to the situation when using a Rb vapor cell. Our analysis also shows that EIA resonances can be observed when the uppermost state is a Rydberg state, which forms an important class of atoms for many fundamental experiments.

\section*{Acknowledgments}
This work was supported by the Department of Science and Technology, India. VB acknowledges financial support from a DS Kothari post-doctoral fellowship of the University Grants Commission, India.

\clearpage
\section*{Appendix 1}
The coupled density matrix equations are calculated by using the definition of the Hamiltonion, density matrix and by  incorporation the decay of each level and repopulation from excited levels. The time-evolution of system for moving atoms is given by
\begin{equation*}
\begin{aligned}
\dot{\rho}_{11}&=\Gamma_{2}\rho_{22}+\Gamma_{3}\rho_{33}+\frac{i}{2}\left(\Omega_{p}^{*}\rho_{12}-\Omega_{p}\rho_{21}\right)
\\&+\frac{i}{2}\left(\Omega_{c1}^{*}\rho_{13}-\Omega_{c1}\rho_{31}\right)\\
\dot{\rho}_{22}&=-\Gamma_{2}\rho_{22}+\Gamma_{4}\rho_{44}+\frac{i}{2}\left(\Omega_{p}\rho_{21}-\Omega_{p}^{*}\rho_{12}\right)
\\&+\frac{i}{2}\left(\Omega_{c2}^{*}\rho_{24}-\Omega_{c2}\rho_{42}\right)\\
\dot{\rho}_{33}&=-\Gamma_{3}\rho_{33}+\frac{i}{2}\left(\Omega_{c1}\rho_{31}-\Omega_{c1}^{*}\rho_{13}\right)\\
\dot{\rho}_{44}&=-\Gamma_{4}\rho_{44}+\frac{i}{2}\left(\Omega_{c2}\rho_{42}-\Omega_{c2}^{*}\rho_{24}\right)\\
\dot{\rho}_{12}&=-\frac{\Gamma_{2}}{2}\rho_{12}+i(\Delta_{p}\pm k_{p}v)\rho_{12}\\
&+\frac{i}{2}\Omega_{p}\left(\rho_{11}-\rho_{22}\right)
-\frac{i}{2}\left(\Omega_{c1}\rho_{32}-\Omega_{c2}^{*}\rho_{14}\right)\\
\dot{\rho}_{13}&=-\frac{\Gamma_{3}}{2}\rho_{13}+i(\Delta_{c1}\pm k_{c1}v)\rho_{13}
\\&+\frac{i}{2}\Omega_{c1}\left(\rho_{11}-\rho_{33}\right)
-\frac{i}{2}\Omega_{p}\rho_{23}\\
\dot{\rho}_{14}&=-\frac{\Gamma_{4}}{2}\rho_{14}+i(\Delta_{p}\pm k_{p}v+\Delta_{c2}\mp k_{c2}v)\rho_{14}
\\&+\frac{i}{2}\left(\Omega_{c2}\rho_{12}-\Omega_{p}\rho_{24}-\Omega_{c1}\rho_{34}\right)\\
\dot{\rho}_{23}&=-\frac{\Gamma_{2}+\Gamma_{3}}{2}\rho_{23}+i(\Delta_{c1}\pm k_{c1}v-\Delta_{p}\mp k_{p}v)\rho_{23}
\\&+\frac{i}{2}\left(\Omega_{c1}\rho_{21}-\Omega_{p}^{*}\rho_{13}-\Omega_{c2}\rho_{43}\right)\\
\dot{\rho}_{24}&=-\frac{\Gamma_{2}+\Gamma_{4}}{2}\rho_{24}+i(\Delta_{c2}\mp k_{c2}v)\rho_{24}
\\&+\frac{i}{2}\Omega_{c2}\left(\rho_{22}-\rho_{44}\right)
-\frac{i}{2}\Omega_{p}^{*}\rho_{14}\\
\dot{\rho}_{34}&=-\frac{\Gamma_{3}+\Gamma_{4}}{2}\rho_{34}\\
&+i(\Delta_{p}\pm k_{p}v+\Delta_{c2}\mp k_{c2}v-\Delta_{c1}\mp k_{c1}v)\rho_{34}
\\&+\frac{i}{2}\left(\Omega_{c2}\rho_{32}-\Omega_{c1}^{*}\rho_{14}\right)\\
\label{equations}
\end{aligned}
\end{equation*}

\section*{Appendix 2}
\begin{figure}[tbh!]
	\centering
	\includegraphics[width=0.5\textwidth]{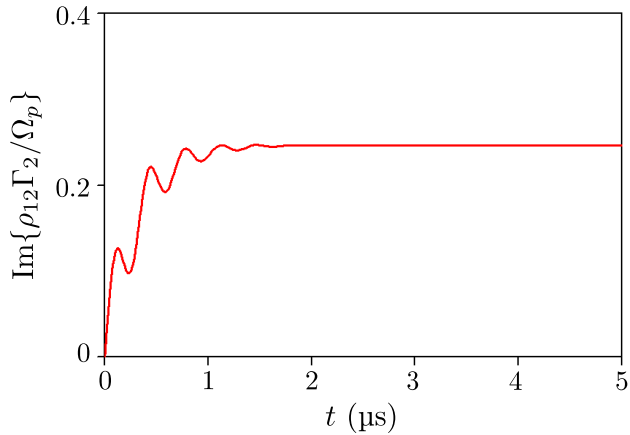}
\caption{(Color online) Time evolution of probe absorption given by Im\{$\rho_{12}\Gamma_{2}/\Omega_{p}$\} for near-perfect match case.}
	\label{Time_evolution}
\end{figure}

In order to verify that the system reaches steady state within a reasonable time, we have studied the transient behavior of the 4-level system. Our calculations show that transients die out after a few microseconds, corresponding to a few lifetimes of the excited level $\ket{2}$. As an example, we show in Fig.\ \ref{Time_evolution} the transient behavior for the case of near-perfect match, where $\ket{2}$ is the $\rm 5P_{3/2}$ state of Rb with a lifetime of 26 ns. In this case, steady state is reached after $ t = 2 $ \textmu s.

\section*{References}

\end{document}